\documentclass[hyper]{JHEP3} 

\usepackage{epsfig}

\newcount\hour
\newcount\minute
\newtoks\amorpm
\hour=\time\divide\hour by60
\minute=\time{\multiply\hour by60 \global\advance\minute by-
\hour}
\edef\standardtime{{\ifnum\hour<12 \global\amorpm={am}%
    \else\global\amorpm={pm}\advance\hour by-12 \fi
    \ifnum\hour=0 \hour=12 \fi
    \number\hour:\ifnum\minute<100\fi\number\minute\the\amorpm}}
\edef\militarytime{\number\hour:\ifnum\minute<100\fi\number\minute}
\def\bold#1{\setbox0=\hbox{$#1$}%
     \kern-.025em\copy0\kern-\wd0
     \kern.05em\copy0\kern-\wd0
     \kern-.025em\raise.0433em\box0 }

\newcommand{\newc}{\newcommand}
\newc\eg{{\it {e.g.}}}  \newc\etal{{\it {et al.}}} \newc\ie{{\it i.e.}}
\newc\etc{{\it {etc}}}  

\newcommand\lsim{\mathrel{\rlap{\lower4pt\hbox{\hskip1pt$\sim$}}
    \raise1pt\hbox{$<$}}}
\newcommand\gsim{\mathrel{\rlap{\lower4pt\hbox{\hskip1pt$\sim$}}
    \raise1pt\hbox{$>$}}}

\newc{\epem}{e^{+}e^{-}}
\newc{\mhalf}{m_{1/2}}      \newc{\mzero}{m_0}

\newc{\tanb}{\tan\beta}
\newc{\azero}{A_0}
\newc{\at}{A_t} \newc{\ab}{A_b} \newc{\atau}{A_\tau} 

\newc{\bmu}{B\mu}           \newc{\sgn}{{\rm sgn}}
\newc{\mone}{M_1}           \newc{\mtwo}{M_2}

\newc{\charone}{\chi_1^\pm} \newc{\mcharone}{m_{\chi_1^\pm}}

\newc{\hl}{h}               \newc{\mhl}{m_{\hl}}
\newc{\hh}{H}               \newc{\mhh}{m_{\hh}}
\newc{\ha}{A}               \newc{\mha}{m_{\ha}}
\newc{\hc}{H^{\pm}}         \newc{\mhc}{m_{\hc}}

\newc{\qzero}{Q_0}          \newc{\qstop}{Q_{\widetilde t}}

\newc{\amu}{a_{\mu}}        \newc{\amususy}{a_{\mu}^{\rm SUSY}}
\newc{\amuexpt}{a_{\mu}^{\rm expt}}        \newc{\amusm}{a_{\mu}^{\rm SM}}
\newc{\deltaamususy}{\Delta a_{\mu}^{\rm SUSY}}
\newc\gmtwo{(g-2)_{\mu}} \newc\deltaamu{\Delta a_{\mu}}

\newc{\msbar}{\overline {\rm MS}} \newc{\drbar}{\overline {\rm DR}}

\newc{\mt}{m_t} \newc{\mb}{m_b} \newc{\mtau}{m_{\tau}}
\newc{\yt}{h_t} \newc{\yb}{h_b} \newc{\ytau}{h_{\tau}}
\newc{\mtpole}{m_t^{\rm pole}} \newc{\mbpole}{m_b^{\rm pole}} 
\newc{\mtaupole}{m_{\tau}^{\rm pole}} 

\newc{\mtmtsmmsbar}{m_t(m_t)^{\msbar}_{{\rm SM}}}
\newc{\mtmtsmdrbar}{m_t(m_t)^{\drbar}_{{\rm SM}}}
\newc{\mtmtmssmdrbar}{m_t(m_t)^{\drbar}_{{\rm SUSY}}}

\newc{\mbmbsmmsbar}{m_b(m_b)^{\msbar}_{{\rm SM}}}
\newc{\mbmzsmmsbar}{m_b(\mz)^{\msbar}_{{\rm SM}}}
\newc{\mbmzsmdrbar}{m_b(\mz)^{\drbar}_{{\rm SM}}}
\newc{\mbmzmssmdrbar}{m_b(\mz)^{\drbar}_{{\rm SUSY}}}
\newc{\mtaumzsmmsbar}{m_{\tau}(\mz)^{\msbar}_{{\rm SM}}}
\newc{\mtaumzsmdrbar}{m_{\tau}(\mz)^{\drbar}_{{\rm SM}}}
\newc{\mtaumzmssmdrbar}{m_{\tau}(\mz)^{\drbar}_{{\rm SUSY}}}

\newc{\mgut}{M_{\rm GUT}}

\newc{\mplanck}{M_{\rm P}}      \newc{\mpl}{M_{\rm Pl}}
\newc{\msusy}{M_{\rm SUSY}}      \newc{\ms}{M_{\rm S}}

\newc{\jxf}{J({\xf})}
\newc{\jxfexact}{J_{\rm exact}({\xf})}  \newc{\jxfexp}{J_{\rm exp}({\xf})}
\newc{\VEV}[1]{\langle #1 \rangle}

\newc{\xf}{x_f}
\newc\vrel{v_{\rm rel}}
\newcommand\mchi{m_{\chi}}              

\newc\sell{{\widetilde e}_L}      \newc\msell{m_{\sell}}
\newc\selr{{\widetilde e}_R}      \newc\mselr{m_{\selr}}

\newc\snue{{\widetilde \nu}_e}      \newc\msnue{m_{\snue}}
\newc\snutau{{\widetilde \nu}_\tau}      \newc\msnutau{m_{\snutau}}

\newc\supl{{\widetilde u}_L}      \newc\msupl{m_{\supl}}
\newc\supr{{\widetilde u}_R}      \newc\msupr{m_{\supr}}
\newc\sdl{{\widetilde d}_L}      \newc\msdl{m_{\sdl}}
\newc\sdr{{\widetilde d}_R}      \newc\msdr{m_{\sdr}}

\newcommand\squark{\widetilde q}        
	\newcommand\msquark{\widetilde{m}_{\squark}}
\newcommand\slepton{\widetilde l}        
	\newcommand\mslepton{\widetilde{m}_{\slepton}}

\newcommand\mgluino{m_{\widetilde g}}

\newc\hpm{H^\pm} \newc\hp{H^+} \newc\hm{H^-} 
\newc\sfermion{\tilde f}  \newc\msfermion{m_{\sfermion}}  

\newc\second{{\rm sec}} 

\newc\alphas{\alpha_s}

\newc\alphaem{\alpha_{em}}

\newc{\gstar}{g_\ast}           \newc{\gsstar}{g_{s\ast}}
\newc{\geff}{g_{\rm eff}}

\newcommand\mz{m_{Z}}

\newc{\sthw}{\sin\theta_W}              \newc{\cthw}{\cos\theta_W}
\newc{\bino}{\widetilde B}              \newc{\wino}{\widetilde W_3^0}
\newc{\higgsinob}{{\widetilde H}^0_b}   \newc{\higgsinot}{{\widetilde H}^0_t}

\newc{\abund}{\Omega h^2}
\newc{\abundchi}{\Omega_\chi h^2}
\newc{\abundcdm}{\Omega_{{\rm CDM}} h^2}
\newc{\omegam}{\Omega_{{\rm M}}}       \newc{\abundm}{\Omega_{{\rm M}} h^2}
\newc{\omegab}{\Omega_{{\rm b}}}        \newc{\abundb}{\Omega_{{\rm b}} h^2}
\newc{\omegacdm}{\Omega_{{\rm CDM}}}   \newc{\omegatot}{\Omega_{{\rm TOT}}}

\newc{\rhocrit}{\rho_{crit}}
\newc{\rhochi}{\rho_{\chi}}

\newcommand\tev{\,\mbox{TeV}}
\newcommand\gev{\,\mbox{GeV}}

\newcommand\pb{\,\mbox{pb}}

\newc\br{\mbox{BR}}

\newc{\ra}{\rightarrow}
\newc{\beq}{\begin{equation}}
\newc{\eeq}{\end{equation}}
\newc{\bea}{\begin{eqnarray}}
\newc{\eea}{\end{eqnarray}}

\renewcommand\]{\right]}

\newcommand\vs{{\it {vs.}}}

\newc\stoponetwo{{\widetilde t}_{1,2}}
\newc\sbotonetwo{{\widetilde b}_{1,2}}
\newc\stauonetwo{{\widetilde \tau}_{1,2}}

\newc\bsgamma{b\ra s \gamma }
\newc\brbsgamma{\br( B\ra X_s \gamma )}

\newc{\sigsip}{\sigma^{SI}_{p}}	\newc{\sigsin}{\sigma^{SI}_{n}}
\newc{\sigsdp}{\sigma^{SD}_{p}}	\newc{\sigsdn}{\sigma^{SD}_{n}}
\newc{\sigsiA}{\sigma^{SI}_{A}}	

\renewcommand\]{\right]}

\long\def\begincomment#1\endcomment{%
        \begingroup\sf\baselineskip12pt#1\endgroup}

\title{Upper and Lower Limits on Neutralino WIMP Mass\\ and
Spin--Independent Scattering Cross Section,\\
and Impact of New \boldmath{$(g-2)_{\mu}$} Measurement
}

\author{Yeong Gyun Kim\\
        Department of Physics, Lancaster University,
        Lancaster LA1 4YB, England\\
        E-mail: \email{Y.G.Kim@lancaster.ac.uk}}

\author{Takeshi Nihei\\
        Department of Physics, College of Science and Technology,
        Nihon University, \\
        1-8-14, Kanda-Surugadai, Chiyoda-ku, Tokyo, 101-8308, Japan \\
        E-mail: \email{nihei@phys.cst.nihon-u.ac.jp}}

\author{Leszek Roszkowski\\
        Department of Physics, Lancaster University,
        Lancaster LA1 4YB, England\\
        E-mail: \email{L.Roszkowski@lancaster.ac.uk}}

\author{Roberto Ruiz de Austri\\
        Physics Division, School of Technology, 
        Aristotle University of Thessaloniki, \\
        GR - 540 06 Thessaloniki, Greece \\
        E-mail: \email{rruiz@gen.auth.gr}}

\abstract{We derive the allowed ranges of the spin--independent
interaction cross section $\sigsip$ for the elastic scattering of
neutralinos on proton for wide ranges of parameters of
the general Minimal Supersymmetric Standard Model.  We investigate the
effects of the lower limits on Higgs and superpartner masses from
colliders, as well as the impact of constraints from $\bsgamma$ and
the new measurement of $\gmtwo$ on the upper and lower limits on
$\sigsip$. We further explore the impact of the neutralino
relic density, including coannihilation, and of theoretical
assumptions about the largest allowed values of the supersymmetric
parameters. For $\mu>0$, requiring the latter to lie below
$1\tev$ leads to $\sigsip\gsim 10^{-11}\pb$ at
$\mchi\sim100\gev$ and $\sigsip\gsim 10^{-8}\pb$ at
$\mchi\sim1\tev$. When the supersymmetric parameters are allowed above
$1\tev$, for $440\gev \lsim \mchi\lsim 1020 \gev$ we derive a {\em
parameter--independent lower limit} of $\sigsip\gsim 2\times
10^{-12}\pb$. (No similar lower limits can be set for $\mu<0$ nor for
$1020\gev\lsim\mchi\lsim2.6\tev$.) Requiring
$\abundchi<0.3$ implies a {\em parameter--independent upper limit}
$\mchi\lsim2.6\tev$. The new $\epem$--based measurement of
$(g-2)_{\mu}$ restricts $\mchi\lsim 350\gev$ at $1\,\sigma$~CL and
$\mchi\lsim515\gev$ at $2\,\sigma$~CL, and implies $\mu>0$.  The
largest allowed values of $\sigsip$ have already become accessible to
recent experimental searches.  }


\begin{document}

\section{Introduction}\label{intro:sec}
The hypothesis of the lightest neutralino $\chi$, as the lightest
supersymmetric particle (LSP), providing the dominant contribution to
cold dark matter (CDM) in the Universe, has inspired much activity in
the overlap of today's particle physics and cosmology. It is
well--known that the relic density of the neutralinos is often
comparable with the critical density~\cite{ehnos84,jkg96}.  The
expectation that the Galactic dark matter (DM) halo is mostly made of
weakly--interacting massive particles (WIMPs) has further led to much
experimental activity. In particular, the experiments looking for CDM
WIMPs elastically scattering off underground targets have recently set
limits on spin--independent (SI), or scalar, cross section of
the order of
$10^{-6}\pb$~\cite{cdms00,edelweiss-june02,ukdmc-zeplini-sep02}.  They
have also nearly ruled out the region of ($\mchi,\sigsip$) that has
been claimed by the DAMA experiment to be consistent with an annual
modulation effect~\cite{dama00}.  Initial and early
studies~\cite{gw85,griest89,ddstudies:past,dn93scatt:ref} were
followed by more recent work~\cite{ddstudies:recent}, where it was
concluded that current experimental sensitivity is generally
comparable with the ranges expected from the neutralino WIMP in the
Minimal Supersymmetric Standard Model (MSSM). It is, however, still at
least on order of magnitude, or so, above the ranges predicted by
recent analyses of the Constrained MSSM
(CMSSM)~\cite{efo00,cmssm:recent, rrn1,lr-latalk,cn0208}.

For comparison, cross sections for spin--dependent (SD) interactions
are in the case of the neutralino generally some two or three orders of
magnitude larger than the SI ones. On the other hand, at present
detectors are still not sensitive enough to explore the parameter
space of the MSSM, despite recent
progress~\cite{ddspin:recentexpt}.
 
In light of the ongoing and planned experimental activities, it is
timely to conduct a thorough and careful re--analysis of the predicted
cross sections for SI scattering of neutralino WIMPs. Such a study is
rather challenging because resulting ranges often strongly depend on a
given SUSY model and on related theoretical assumptions. They are
further affected by experimental limits on SUSY, both from colliders
and from indirect searches, as well as by cosmological input, where
the relic abundance of the CDM has been measured with better accuracy
both directly and in CMBR studies~\cite{cmbstudiesrecent}.  Over the
last few years and months there have been also new results for LEP
lower bounds on the masses of the lightest Higgs and
electroweakly--interacting superpartners, Tevatron lower limits on
strongly--interacting superpartners, as well as limits on allowed SUSY
contributions to $\bsgamma$, and especially to the anomalous magnetic
moment of the muon $\gmtwo$~\cite{gm202}. The new result for $\gmtwo$
indicates a sizable deviation from the Standard Model prediction,
whose value is still a subject of much discussion.  As we will show,
when interpreted in terms of SUSY, the required extra contribution to
$\gmtwo$ plays a unique role in implying a stringent {\em upper} bound
$\mchi\lsim 350\gev$ ($1\,\sigma$~CL) and $\mchi\lsim
512\gev$ ($2\,\sigma$), but it does {\em not} affect much the
allowed ranges of the SI scattering cross section $\sigsip$.

In this paper we carefully study the impact of the above
constraints. In an attempt to minimize theoretical bias, we work here
in the context of the general MSSM, which will be defined below, with
an additional assumption of $R$--parity conservation. We focus here on
the SI cross section case. Other recent studies of the general MSSM
include~\cite{efo01:gmssm,mpmg00,bg02:mssm}.  The results presented
here show that the level of experimental sensitivity that has recently
been reached~\cite{cdms00,edelweiss-june02,ukdmc-zeplini-sep02} has
now indeed allowed one to start exploring cosmologically favored
ranges of the neutralino WIMP mass and SI cross section. However, we
point out a number of caveats and relations and further discuss the
origin of, and robustness of, the upper and lower limits on
$\sigsip$. In particular, for a big range of the heavy neutralino mass
$440\gev\lsim\mchi\lsim 1020\gev \gev$ we are able to derive {\em
parameter--independent} lower bounds on $\sigsip$. Collider lower
limits on Higgs and superpartner masses plus requiring $\abundchi<0.3$
alone leads to a {\em parameter--independent} upper bound
$\mchi\lsim2.6\tev$.

\section{The MSSM}\label{mssm:sec}
We start by reviewing relevant features of the general
MSSM~\cite{susyreview}. (We follow the convention of~\cite{gh}.)
By the MSSM we mean a supersymmetrized version of the Standard Model,
with Yukawa and soft SUSY--breaking terms consistent with
$R$--parity. We neglect CP--violating phases in the Higgs and SUSY
sectors and assume no mixings among different generations of squarks
and sleptons, since both are probably small.
In the MSSM, all the slepton and squark masses can be
considered as a priori free
parameters set at the electroweak scale. In the same way we treat the
trilinear parameters $A_i$ ($i=t,b,\tau$) of the third generation
while neglecting the ones of the first two.
The Higgs sector is determined at the tree level by the usual ratio
of the neutral Higgs VEV's $\tanb=v_t/v_b$ and the mass of the
pseudoscalar $\mha$. In computing full one-loop and leading two-loop
radiative corrections to the lightest scalar Higgs we use the package
FeynHiggsFast (FHF)~\cite{feynhiggsfast:ref}. As we will see,
Higgs masses will play an important role in the analysis.

The lightest neutralino $\chi$ is lightest of the four mass eigenstates
of the linear combinations of the bino $\bino$, the wino $\wino$ and the two
higgsinos $\higgsinob$ and $\higgsinot$
\beq
\chi\equiv\chi^0_1= 
N_{11}\bino + N_{12}\wino + N_{13}\higgsinob + N_{14}\higgsinot.
\label{chidef:eq} 
\eeq

The neutralino mass matrix ${\cal M}$~\cite{gh} is determined by
the $U(1)_Y$ and $SU(2)_L$ gaugino mass parameters $\mone$
and $\mtwo$, respectively (and we impose the relation
$\mone=\frac{5}{3}\tan^2\theta_W  \mtwo $, which comes from assuming gaugino
mass unification at GUT scale),
the Higgs/higgsino mass parameter $\mu$, as well as $\tanb$.
In the region $|\mu|\gg\mone$, the lightest 
neutralino is mostly a bino with mass
$\mchi\simeq\mone$. In the other extreme, it is mostly
higgsino--dominated and $\mchi\simeq|\mu|$.



For the purpose of this analysis, we take as independent parameters:
$\tanb$, $\mu$, $\mtwo$, $\mha$, $A_{t,b}$ as well as the soft masses
of the sleptons and of the squarks. In order to make our analysis
manageable, we make an additional assumption that, at the electroweak
scale, the soft mass parameters of the sleptons are all equal to some
common value $\mslepton$, and analogously $\msquark$ for all the
squarks.  One normally expects certain relations among the physical
masses of the sleptons and the squarks since they in addition receive
well--defined D--term and F--term contributions to their mass
matrices. Assuming common soft mass terms, at either GUT or
electroweak scale, this normally leads to the sleptons being lighter
than the squarks. Furthermore, for the sfermions of the 3rd generation
it is natural to expect large mass splittings.  However, we believe
that, for our purpose, introducing just two separate common soft mass
scales $\mslepton$ and $\msquark$, while greatly simplifying the
analysis, will not play much role in our overall conclusions for the
SI cross sections. (This is in contrast to often assumed {\em full}
degeneracy of soft sfermion masses which leads in our opinion to an
unnecessary limitation on the allowed SUSY parameter space.)

What we do find important is to disentangle squark and slepton masses.
This is mainly because experimental limits on slepton masses are
significantly weaker than in the case of squarks. For the case of the 
bino--like neutralino, which is the most natural case for providing
$\abundchi\sim1$~\cite{chiasdm}, it is therefore the mass of the
lightest slepton which often predominantly determines
$\abundchi$. Assuming common soft masses for sleptons and squarks at the
electroweak scale would therefore generally lead to overestimating
$\abundchi$ by missing the cases where relatively light sleptons
(below current squark mass bounds) would otherwise provide acceptable
$\abundchi$. An additional effect is that of coannihilation.  When
slepton mass is only somewhat larger than that of the LSP, the
neutralino relic abundance is strongly reduced and otherwise forbidden
cases become allowed~\cite{sleptoncoann:orig}.

\section{The Spin--Independent Cross Section}\label{sics:sec}
For non-relativistic Majorana particles, like the neutralino WIMP, the
elastic scattering off constituent quarks and gluons of some nucleon
$^A_Z X$ is given by an effective differential cross
section~\cite{gw85,griest89,dn93scatt:ref} 
\begin{eqnarray}
\frac{d\sigma}{d|\vec{q}|^2}=\frac{d\sigma^{SI}}{d|\vec{q}|^2}+
\frac{\ d\sigma^{SD\ }}{d|\vec{q}|^2},
\label{signucleus:eq}
\end{eqnarray}
where the transferred momentum $\vec{q}=\mu_A\vec{v}$ depends on the
velocity $\vec{v}$ of the incident WIMP, and $\mu_A=m_A \mchi/(m_A+\mchi)$ 
is the reduced mass of the system.  The effective
WIMP-nucleon cross sections $\sigma^{SI}$ and $\sigma^{SD}$ are
computed by evaluating nucleonic matrix elements of corresponding
WIMP--quark and WIMP--gluon interaction operators.

In the SI part, contributions from individual
nucleons in the nucleus add coherently and the finite size effects are
accounted for by including the SI nuclear form factor $F(q)$.
The differential cross section for the scalar part then takes the form 
\cite{jkg96}     
\begin{eqnarray}
\frac{d\sigma^{SI}}{d|\vec{q}|^2}=\frac{1}{\pi v^2}\[Z f_p +(A-Z) f_n\]^2
F^2 (q),
\end{eqnarray}
where $f_{p}$ and $f_{n}$ are the effective neutralino couplings to
protons and neutrons, respectively. Explicit expressions for the case
of the supersymmetric neutralino can be found, \eg, in~\cite{bb98}. 
The formalism we follow has been reviewed in several recent
papers~\cite{jkg96,bb98,efo00}. We have 
re--done the original complete calculation of Drees and
Nojiri~\cite{dn93scatt:ref} and agreed with their results. 

A convenient quantity which is customarily used in comparing theory
and experimental results for SI interactions is the
cross section  $\sigsip$ 
for WIMP elastic scattering of free proton in the limit of zero
momentum transfer:
\beq
\sigsip= \frac{4}{\pi}\mu^2_p f_p^2
\label{sigsipdef}
\eeq
where $\mu_p$ is defined similarly to $\mu_A$ above. The
analogous quantity for a target with nuclei with mass number $A$ can then
be expressed in terms of $\sigsip$ as 
\beq
\sigsiA= \frac{4}{\pi}\mu^2_A \[Z f_p +(A-Z) f_n\]^2 =
\left(\frac{\mu_A}{\mu_p}\right)^2 A^2 \sigsip.
\label{sigsiAdef}
\eeq
One can do so because, for Majorana WIMPs, $f_p\simeq f_n$.  

The coefficients $f_{p,n}$ can be expressed as~\cite{dn93scatt:ref}
\bea
{f_p \over m_p} = \sum_{q=u,d,s} {f_{Tq}^{(p)} \over m_q} f_q ~+~ 
{2\over 27} f^{(p)}_{TG} \sum_{c,b,t} {f_q \over m_q} + ... 
\nonumber
\eea
where $f^{(p)}_{TG} = 1 - \sum_{q=u,d,s} f^{(p)}_{Tq} $, and
$f_{Tq}^{(p)}$ 
is given by
$<p|m_q \bar{q} q|p> = m_p f_{Tq}^{(p)}$ ($q=u,d,s$),
and analogously for the neutron. The masses and ratios
$B_q=\langle p|{\bar q}q|p\rangle$ of light constituent quarks in a
nucleon come with some uncertainties. 
For definiteness, we
follow a recent re--evaluation~\cite{efo00} and assume $m_u/m_d=
0.553\pm0.043$, $m_s/m_d= 18.9\pm0.8$, and $B_d/B_u=
0.73\pm0.02$, as well as 
\begin{eqnarray}
f^{(p)}_{Tu}=0.020\pm0.004,~ f^{(p)}_{Td}=0.026\pm0.005,~
f^{(p)}_{Ts}=0.118\pm0.062 
\nonumber \\ \nonumber\\
f^{(n)}_{Tu}=0.014\pm0.003,~ f^{(n)}_{Td}=0.036\pm0.008,~
f^{(n)}_{Ts}=0.118\pm0.062,
\nonumber
\end{eqnarray}
which numerically gives SI cross section values very similar to using
the set of~\cite{jkg96}.  Some other recent studies use a new
determination of $\sigma_{\pi N}$ to derive a much larger value for
$f^{(p)}_{Ts}\simeq0.37$~\cite{hm:except}. We find such values somewhat
questionable since they imply that the strange quark component of the
nucleon would be larger than the up and down ones. We have numerically
checked that using the set of input parameters of~\cite{hm:except}
gives typically SI cross section values a factor of six higher than in
our case.

It is worth noting that, despite several different diagrams and
complicated expressions, it is the exchange of the heavy scalar Higgs
that in most cases comes out to be numerically dominant. It is
further enhanced when the neutralino is a mixed
gaugino--higgsino state~\cite{bfg89}. As
$\mhh\simeq\mha$ increases, $\sigsip$ drops as $\mhh^{-4}$ because of
the $t$--channel propagator effect in $\chi q\ra \chi q$ elastic
scattering. Eventually, at smaller $\sigsip$, squark exchange becomes
important, and even dominant, instead.

\section{Details of the Scan}\label{scan:sec}
As outlined above, we use seven parameters $\tanb$, $\mtwo$,
$\mu$, $\mha$, $\msquark$, $\mslepton$ and $\at=\ab$ to
conduct a careful scan of the general MSSM parameter space. For their
allowed ranges we take:
\bea
50\gev\leq &\mtwo& \leq2\tev \nonumber \\
50\gev\leq &|\mu|&  \leq2\tev~(4\tev)  \nonumber \\
50\gev\leq &\mslepton & \leq 2\tev~(4\tev) \nonumber \\
200\gev\leq &\msquark& \leq 2\tev ~(4\tev) \label{parrange:eq} \\
90\gev\leq & \mha &\leq 2\tev \nonumber \\
0\leq &|A_{t,b}|&\leq 1\tev\nonumber \\
5\leq &\tanb&  \leq 65 \nonumber
\eea
while we set $~A_\tau=0$ since 
we treat the masses of the slepton as independent
parameters anyway. In addition to a general scan of the parameter space, in 
many cases we do several focused scans and explore the
effect of extremely large values of $\mu$, $\mslepton$ and $\msquark$
beyond $2\tev$ (given in brackets above) 
in order to derive {\em parameter--independent} lower
limits on $\sigsip$ for a big range of large $\mchi$, as described below.
The minimum values are set so that the resulting physical masses
of Higgs and superpartners are limited from below by collider
bounds. For the lighter chargino we take
$\mcharone>104\gev$~\cite{lepsusy}, for sleptons the lower limit of
$90\gev$~\cite{lepsusy} and for squarks
$200\gev$~\cite{squarkcdf}. The lower limit on $\mchi$
depends not only on a model but also on a number of additional
assumptions~\cite{lipniackatalk}. For this reason, in our analysis we
conservatively 
do not impose a direct experimental limit on $\mchi$, but instead infer
it from the other limits, especially the one on the chargino mass.  As
regards the lightest Higgs mass $\mhl$, in much of the parameter space
($\mha> 120\gev$) the lower limit on the Standard Model Higgs of
$114.1\gev$~\cite{lephiggs} applies. However
there are two important points to note. Firstly, theoretical
uncertainties in computing $\mhl$ in the MSSM are estimated at
$2-3\gev$. Conservatively, we thus require only $\mhl>
111\gev$. Secondly, for $90\gev<\mha< 120\gev$, there still remains a
sizable range of the ($\mhl,\mha$)--plane where the lightest Higgs
mass given roughly by $\mhl> 0.78 (\mha + 21.7\gev)$ is
allowed~\cite{lephiggs}. We will comment below on the 
effect of this low--mass range on increasing the largest allowed values of
$\sigsip$.

Among indirect limits on SUSY, $b\rightarrow s \gamma$ often places an
important additional constraint on the allowed parameter space. We
calculate the SM contribution to $\br( B\ra X_s \gamma )$ at the full
NLO level and include dominant $\tanb$--enhanced NLO
SUSY~\cite{bsgammasusytanb}, and also include the $c$-quark mass
effect on the SM value~\cite{gm01}. At this level of accuracy the
SM prediction is $\br(B \to X_s\gamma)=(3.70
\pm0.30)\times10^{-4}$~\cite{gm01}. This range is partially
overlapping with the new world--average~\cite{or1}
\beq 
\br(B \to X_s\gamma)=(3.41 \pm0.36)\times10^{-4}
\label{bsgexp02}
\eeq
which has gone up from the previous range of $(3.23\pm 0.72)\times 10^{-4}$
following the new result from BaBAR~\cite{babarbsg02}. 
As described in more detail
in~\cite{rrn1}, with an update in~\cite{moriond02lr}, 
we accordingly allow the full
SM+SUSY contribution to be in the range 
$\br( B\ra X_s \gamma ) = (3.41\pm 0.67)\times 10^{-4}$. 
It is important, however, to stress here an
important salient point.  In computing the SUSY contribution to
$\brbsgamma$ one usually makes an implicit assumption of minimal
flavor violation in the down--type squark sector, which is
theoretically poorly justified. Even a slight modification of the
assumption often leads to a significant relaxation of the bound from
$\bsgamma$ to the point of even allowing $\mu<0$ and relatively light
superpartner masses~\cite{or1}.

A very recent measurement of the anomalous magnetic moment of the muon
$\amu=(g_\mu-2)/2$~\cite{gm202} has confirmed a previous
value~\cite{gm201} but with twice--increased precision~\cite{gm202}
\beq 
\amuexpt-11659000\times10^{-10}= (203\pm8)\times10^{-10}.
\label{amuexpt}
\eeq 
The LO hadronic vacuum polarization contribution has recently been
re--evaluated in~\cite{dehz02,hmnt02} and the light--by--light corrections 
in~\cite{lblupdate}. In~\cite{dehz02} the updated SM prediction
of~\cite{dh98} has
been found to be $\amusm-11659000\times10^{-10}=
(169.1\pm7.8)\times10^{-10}$ when applying data from $\epem$
annihilation cross 
sections and $(186.3\pm7.1)\times10^{-10}$ when applying $\tau$--decay
data. 
This leads to a $3\,\sigma$ discrepancy
\beq
\deltaamu = \amuexpt-\amusm=(33.9\pm11.2)\times10^{-10}
\label{deltaamu:ref}
\eeq 
when using the $\epem$--based data, or to a $1.6\,\sigma$ deviation
\beq
\deltaamu = \amuexpt-\amusm=(16.7\pm10.7)\times10^{-10}
\label{deltaamutau:ref}
\eeq 
when applying the $\tau$--based data. 

If interpretted in terms of SUSY, eq.~(\ref{deltaamu:ref})
restricts the allowed SUSY contribution to
\bea
22.7\times10^{-10}<&\amususy &< 45.1\times10^{-10}~~~(1\,\sigma)\\
\label{amususy1sig}
11.5\times10^{-10}<&\amususy &< 56.3\times10^{-10}~~~(2\,\sigma).
\label{amususy2sig}
\eea
Similar ranges are obtained by using the results
of~\cite{hmnt02,derafael02}. 
The eqs.~(\ref{deltaamu:ref}) and~(\ref{deltaamutau:ref}) further imply 
that $\mu>0$ (the sign of the SUSY contribution is the same as that of
$\mu$).  

This is clearly an intriguing hint for ``new physics''. However, since
some other recent evaluations tend to give a larger value of
$\amusm$~\cite{narisonandty01} and a larger error bar~\cite{rw02},
at this point we will not strictly impose the $\gmtwo$ constraint on
the parameter space that is otherwise allowed by all other
constraints. Nevertheless, below we will discuss the important impact
it has on the upper bounds on $\mchi$.

As regards the WIMP relic abundance, a lower limit on the age of the
Universe conservatively gives $\abundchi<0.3$, while ``direct''
measurements of the CDM lead to $0.1<\abundchi<0.2$ which we will
treat as a preferred range. Recent studies of the CMBR seem to imply
even more restrictive ranges; for example, $\abundchi<0.12\pm0.04$
in~\cite{ms02}.  We will not apply this narrower range yet but will
comment below on its impact on the upper and/or lower limits on
$\sigsip$. We compute $\abundchi$ as accurately as one reasonably can,
at the level of a few per cent both near and further away from poles
and thresholds, by applying our recently derived exact analytic
expressions for neutralino pair--annihilation~\cite{nrr1+2} and
neutralino--slepton coannihilation~\cite{nrr3}, and by using an exact
procedure for the neutralino coannihilation with chargino and
next--to--lightest neutralino~\cite{eg97,darksusy}.

\begin{figure}[t!]
\vspace*{-0.75in}
\hspace*{-.70in}
\begin{center}
\begin{minipage}{3.5in}
\epsfig{file=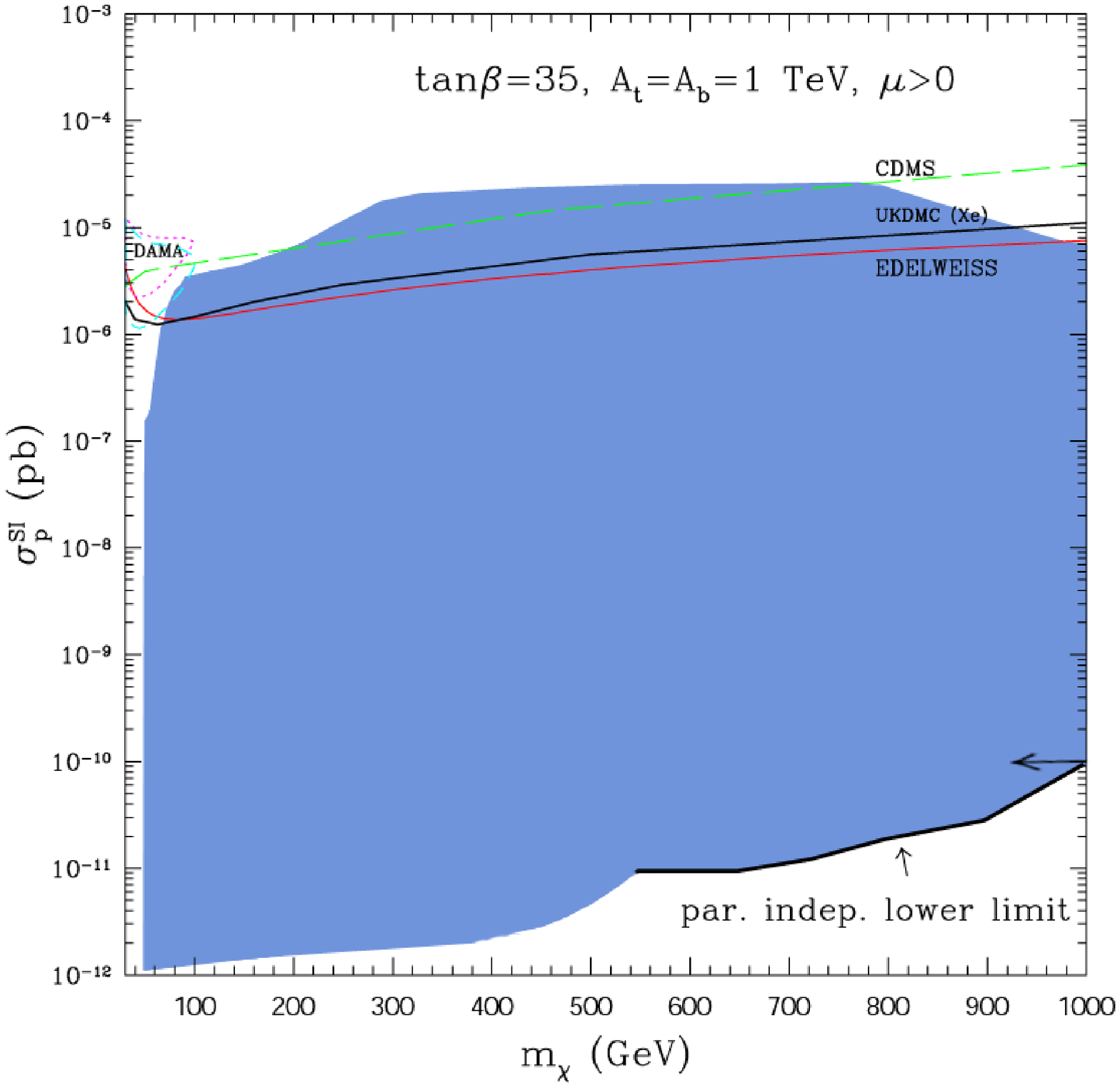,height=3.15in}
\end{minipage}
\caption{\label{tb35a1tall:fig} {\small 
Ranges of $\sigsip$ in the general MSSM
\vs\ $\mchi$ for $\tanb=35$, $\at=\ab=1\tev$ and $\mu>0$, which are allowed by
the bounds from colliders, $\bsgamma$ and $0.1<\abundchi<0.2$, but not 
from $\gmtwo$. Also marked are
some results of recent experimental WIMP searches. The thick black
line (and a left--pointing arrow)
indicates a {\em parameter--independent} lower bound on $\sigsip$ for
$550\gev<\mchi<1020\gev$. No similar bound can be set for lower $\mchi$ because
of the neutralino--slepton coannihilation effect, as explained in the text.
}
}
\end{center}
\end{figure}

\section{Results}\label{results:sec}
The allowed ranges of the SI cross section that result from our scans
are illustrated in Fig.~\ref{tb35a1tall:fig} for $\tanb=35$,
$\at=\ab=1\tev$ and $\mu>0$.  Since $\sigsip$ generally grows with
$\tanb$ due to an enhancement in the heavy scalar Higgs coupling to
down--type quarks, the above choice is a
reasonable compromise between the low and the very large values of $\tanb$.  
In deriving the allowed ranges of $\sigsip$ we have imposed the bounds from
colliders and from $\bsgamma$, and have
further required $0.1<\abundchi<0.2$, as described earlier, but {\em
not} yet the bound from $\gmtwo$. We also mark with a thick solid line
a {\em parameter--independent} lower bound on $\sigsip$ which will be
explained below.

We can see a big spread of $\sigsip$ over some seven (four) orders of
magnitude at small (large) $\mchi$. In fact, the upper values of
$\sigsip$ exceed the latest experimental limits, including the recent
result from Edelweiss~\cite{edelweiss-june02} and the new
limit from the UKDMC Zeplin~I detector~\cite{ukdmc-zeplini-sep02}. 
Also shown is the CDMS bound and the so--called DAMA region. It
is clear that today's experiments have already started probing the
most favored ranges of $\sigsip$ that come from SUSY predictions for
neutralino cold dark matter.

\begin{figure}
\vspace*{-0.75in}
\hspace*{-.70in}
\begin{center}
\begin{minipage}{6.5in}
\epsfig{file=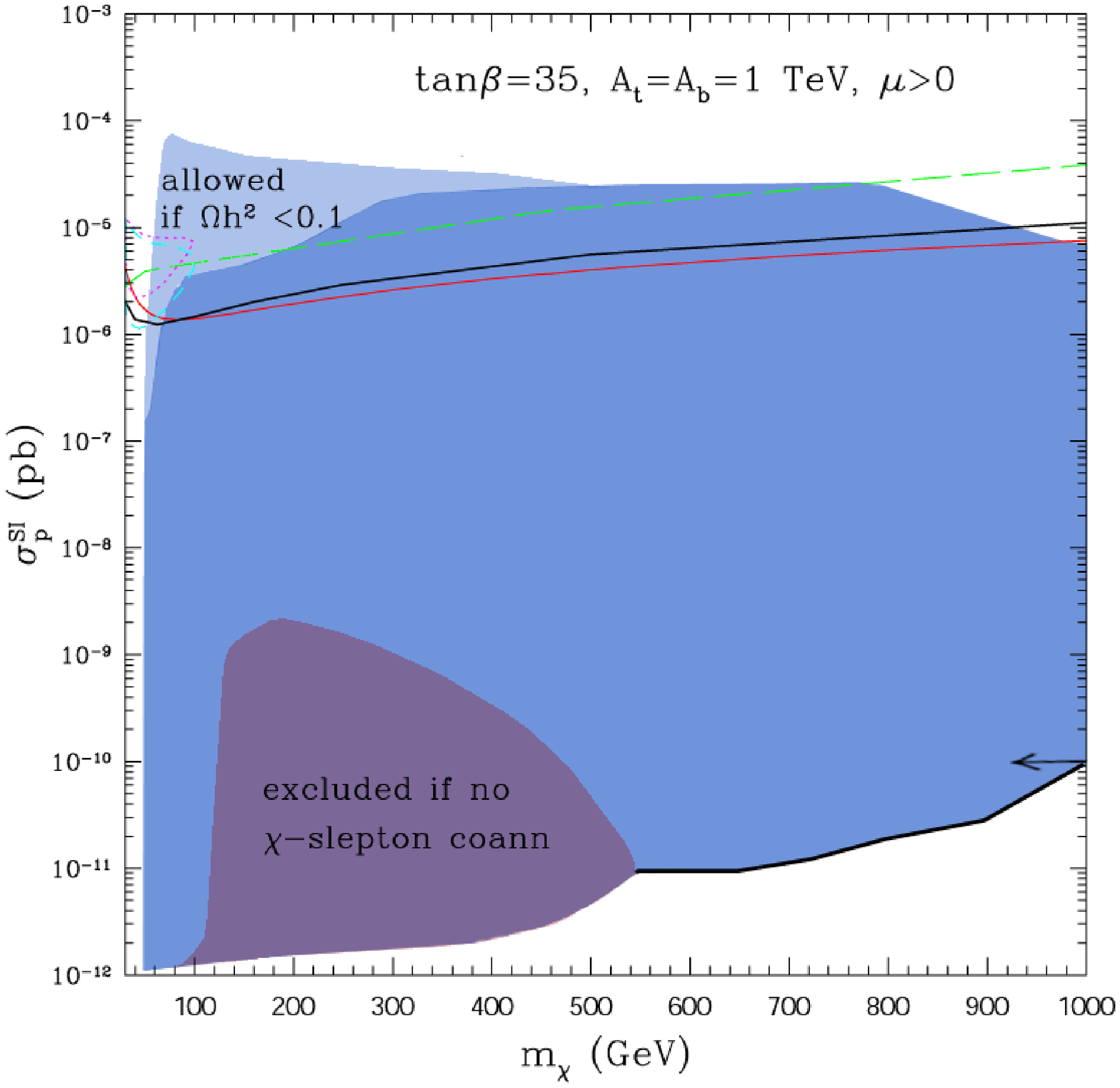,height=3.15in}
\hspace*{-0.18in}
\epsfig{file=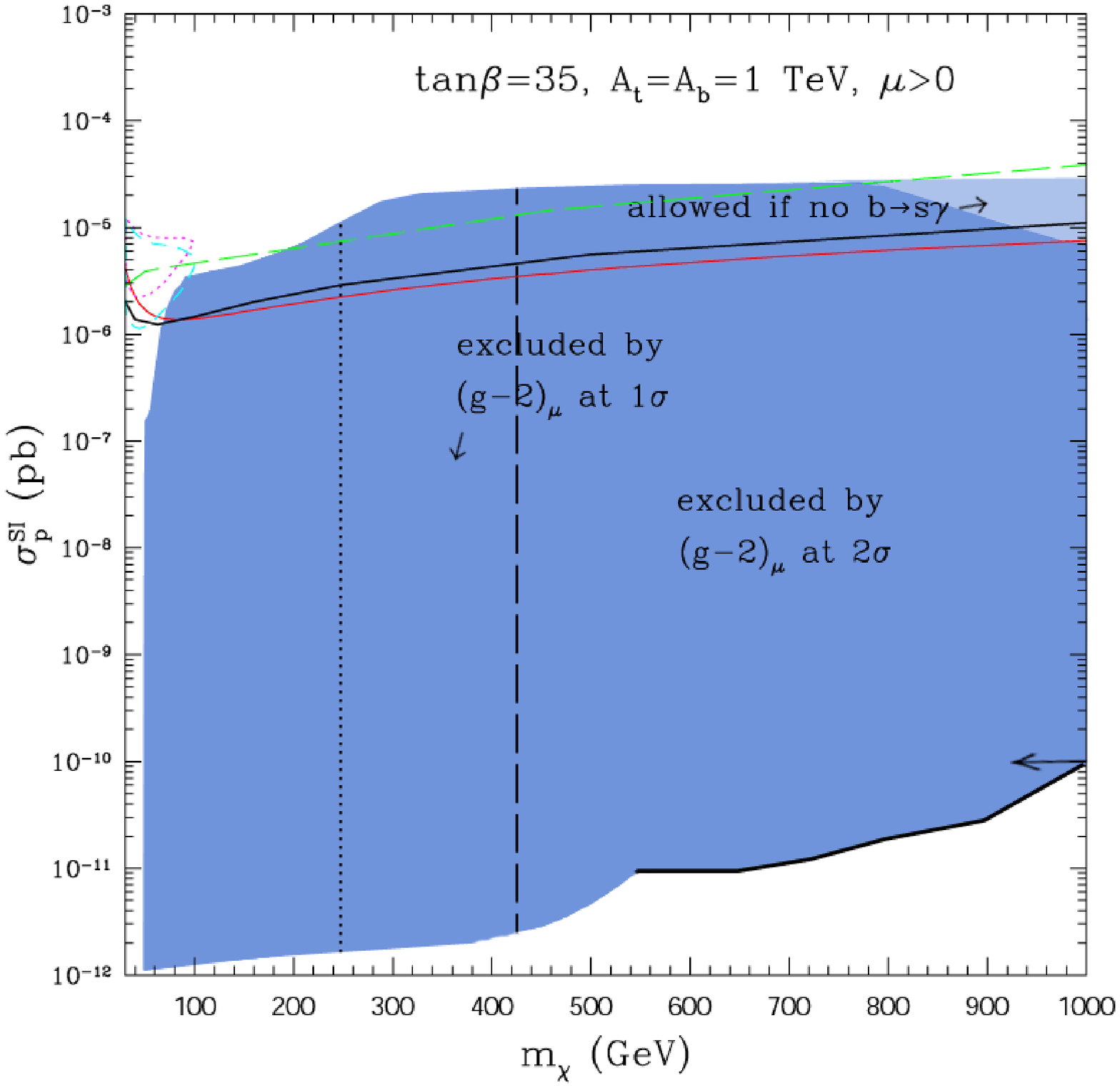,height=3.15in}
\end{minipage}
\end{center}
\vspace*{-.50in} 
\hspace*{-.70in}
\begin{center}
\begin{minipage}{6.5in}
\epsfig{file=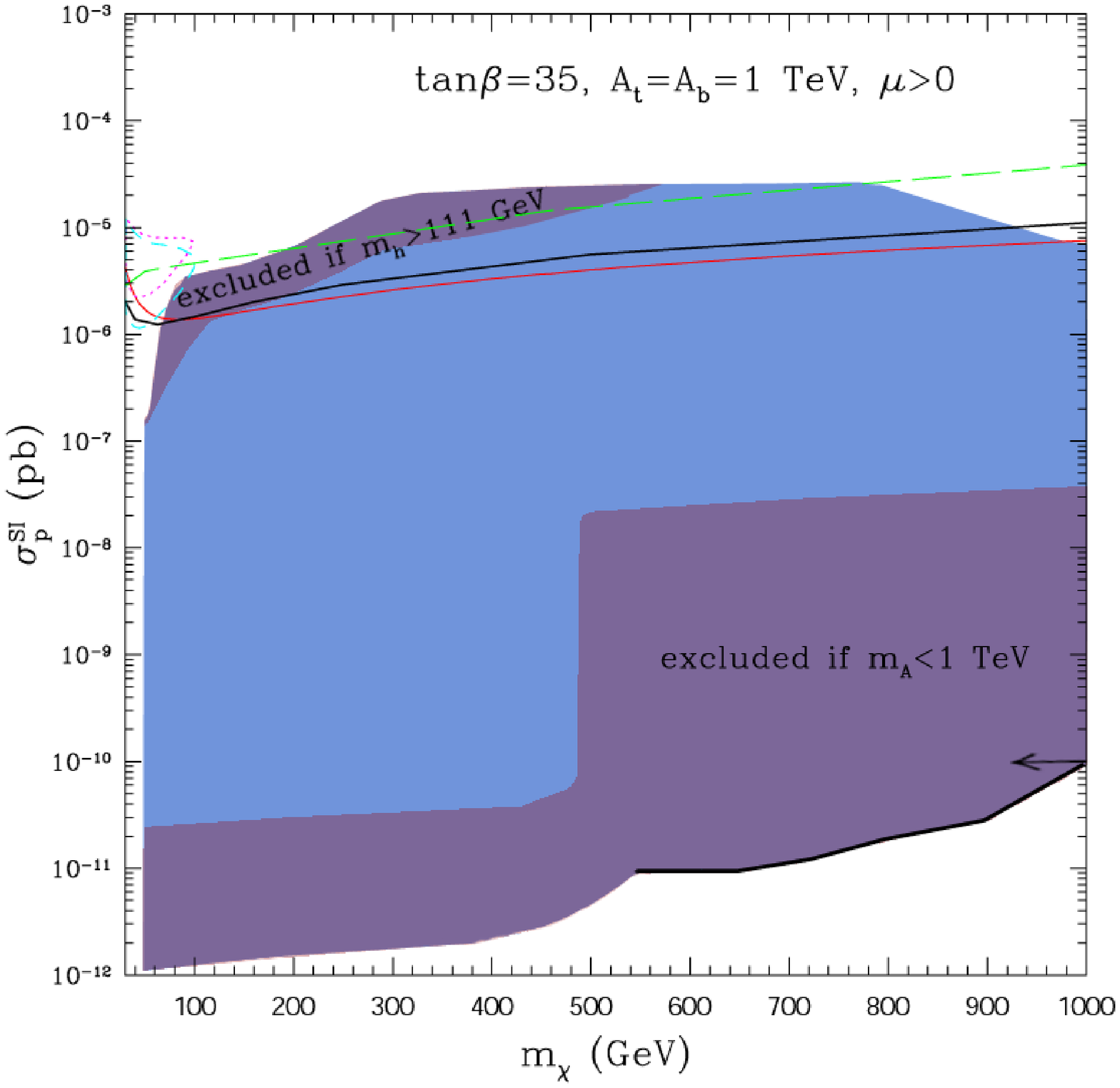,height=3.15in}
\hspace*{-0.18in}
\epsfig{file=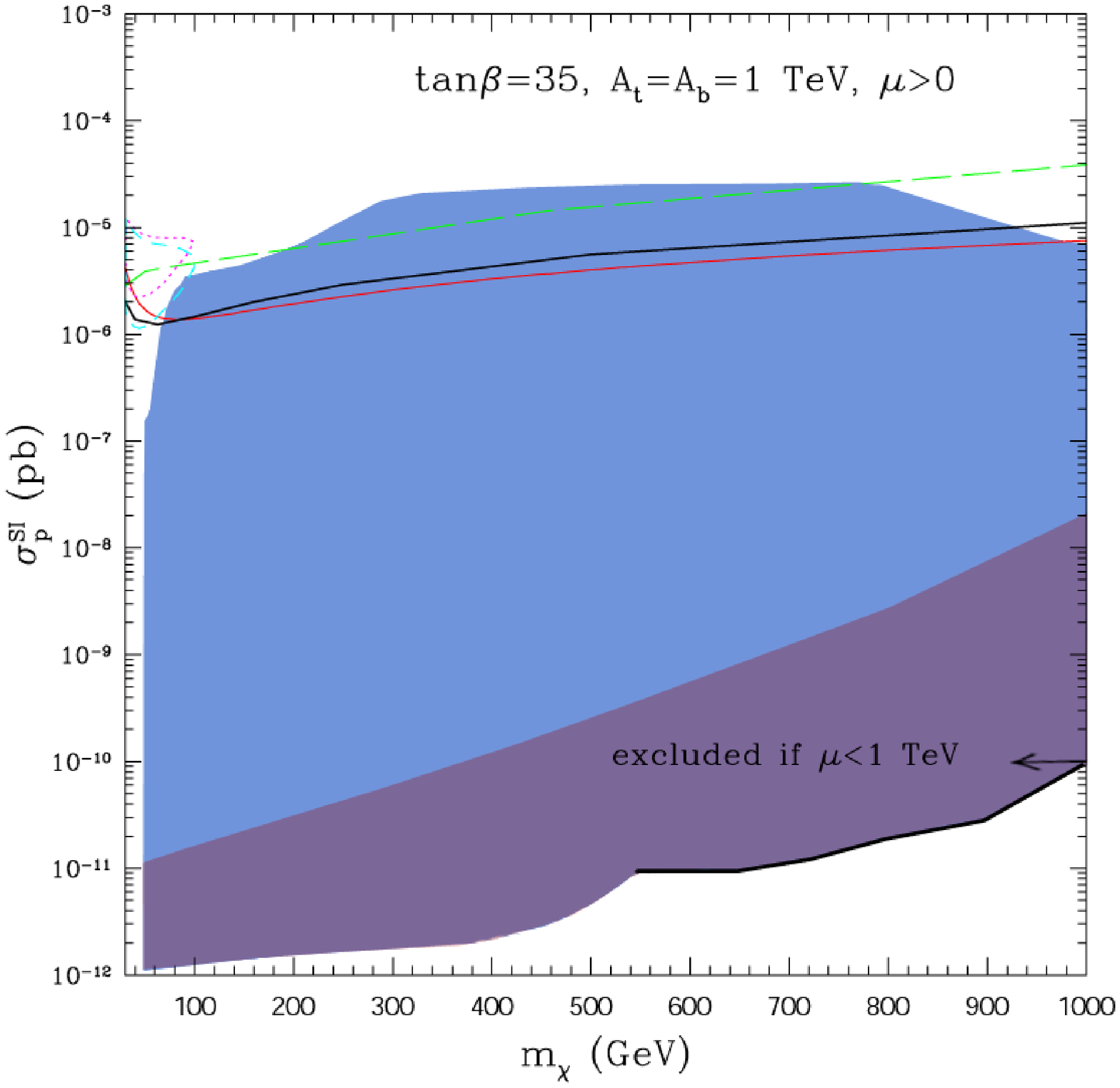,height=3.15in}
\end{minipage}
\caption{\label{tb35a1tdep:fig} {\small Sensitivity of the upper and lower
limits on $\sigsip$ in Fig.~\protect\ref{tb35a1tall:fig} to various
assumptions and constraints. Upper left window: the light blue
region would be allowed if 
$\abundchi<0.1$. The dark--red region would be excluded if
one neglected the effect of neutralino--slepton coannihilation.  
Upper right window: the light blue region would be allowed if one lifted
the constraint from $\bsgamma$. The regions to the right of the
vertical dotted (dashed) lines are excluded by imposing current
$1\,\sigma$ ($2\,\sigma$)~CL bound from $\gmtwo$.
Lower left window: the upper dark--red region would be excluded by assuming
$\mhl>111\gev$ for all $\mha$ (\ie, by neglecting a window of lighter
$\mhl$ which is still allowed for $\mha<120\gev$). Also shown in this
window is the effect of restricting $\mha<1\tev$. 
Lower right window: the same as for $\mha$ but for $\mu$. }}
\end{center}
\end{figure}

In the four windows of Fig.~\ref{tb35a1tdep:fig} we show the effect of
the most important constraints on the upper and lower limits on the
allowed ranges of $\sigsip$. Firstly, in the upper left window we show
the effect of relaxing the cosmological bound by allowing
$\abundchi<0.1$. Obviously, larger ranges of $\sigsip$ now become
allowed since an enhancement in the neutralino pair--annihilation
cross section often, by crossing symmetry, implies an increase in
$\sigsip$. Note, however, that a combination of all the other
constraints, most notably a lower limit on $\mhl$ from LEP and the
constraint from $\bsgamma$, prevents $\sigsip$ from rising by more
than about one order of magnitude and only for not very large values
of $\mchi$. On the other hand, imposing a narrower range
$0.08<\abundchi<0.16$ has almost no effect on the upper and lower
limits on $\sigsip$, although it does remove a number of points from
the allowed ranges of $\sigsip$.

In the same window we also show an important effect, already pointed
out in~\cite{mpmg00}, of including neutralino coannihilation with
sleptons (predominantly with the lighter stau) on allowing very low
ranges of $\sigsip$ at smaller $\mchi$. At lowest $\mchi\lsim120\gev$
the relic abundance can be reduced to the favored range by choosing in
the scan light enough sleptons, even without coannihilation. By
simultaneously choosing large enough heavy Higgs and squark masses,
one can reduce $\sigsip$ to very low values of a few
$\times10^{-12}\pb$. As $\mchi$ increases, $\abundchi$ would normally
increase as well, and become too large, but it is there that
neutralino--slepton coannihilation kicks in.  Since $\sigsip$ is
independent of the slepton masses, by carefully scanning the parameter
space, one can always find $\mslepton$ not much above $\mchi$, in
which case $\abundchi$ can be sufficiently reduced again to fall into
the favored range. The effect is very strong for smaller $\mchi$, thus
explaining a sharp rise of the left side of the dark--red region
allowed by neutralino--slepton coannihilation, but, as the process
becomes increasingly inefficient at larger
$\mchi$~\cite{sleptoncoann:orig,nrr3}, it gradually fades away.

In the upper right window of Fig.~\ref{tb35a1tdep:fig} we present the
effect of imposing the constraint from $\gmtwo$.  We can see that, for
this case, $\mchi\lsim245\gev$~($1\,\sigma$~CL) and
$\mchi\lsim420\gev$~($2\,\sigma$~CL).  This upper limit comes from the
fact that, as $\mchi$ increases, the SUSY contribution from the
$\chi-{\widetilde\mu}$ and $\chi^{-}-{\widetilde\nu}_\mu$ loops become
suppressed and at some point becomes too small to explain the apparent
discrepancy between the SM and the experimental
measurement~\cite{mchifromgmtwo,recentgm2studies}.  On the other hand,
the upper and lower limits on $\sigsip$ are not really affected.

In the same window we also show the effect of relaxing the constraints
from $\bsgamma$. We can see that if it were not imposed, at large
$\mchi$ the upper limit on $\sigsip$ would significantly increase.  In
this region, the mass of the pseudoscalar Higgs, and therefore also
the heavier scalar, is rather small, thus giving larger
$\sigsip$. However, the mass of the charged Higgs is then also on the
lower side, and a cancellation between a charged Higgs--top quark loop
and chargino--stop loop contribution is not sufficient to reduce
$\brbsgamma$ to agree with the experimental limit.

However, we remind the reader that a slight relaxation of the
underlying assumption of minimal flavor violation in the squark sector
often leads to a significant weakening of the bound from $\bsgamma$ to
the point of even allowing  $\mu<0$~\cite{or1}. We would therefore
be cautious in applying the $\bsgamma$ constraint rigidly.

In the lower left window of Fig.~\ref{tb35a1tdep:fig} we present the
sensitivity of the {\em upper} limit on $\sigsip$ to the lower limit
on the light Higgs mass. We can see that it would sizably decrease if
we neglected the region of small $90\gev\lsim\mhl\lsim 111\gev$ which
is still allowed at $\mha\lsim120\gev$, and instead required
$\mhl>111\gev$ for all $\mha$. Note also that the new experimental
limits on $\sigsip$ are for the most part inconsistent with the
possibility of the light Higgs scalar.

\begin{figure}[t!]
\vspace*{-0.75in}
\hspace*{-.70in}
\begin{center}
\begin{minipage}{3in}
\epsfig{file=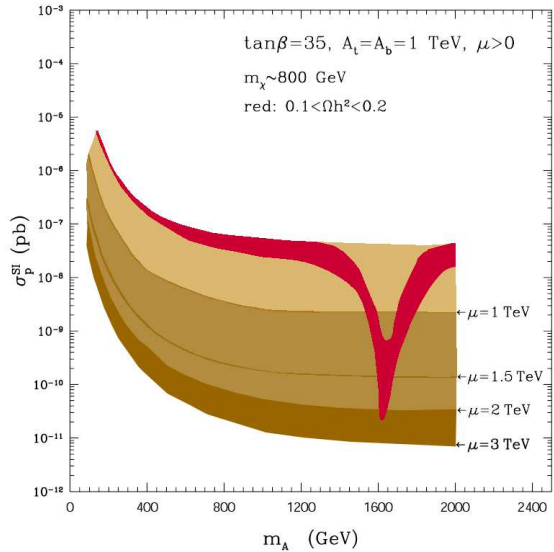,height=3in}
\end{minipage}
\caption{\label{masigtb35a1k:fig} {\small Sensitivity of $\sigsip$ to
$\mha$ and $\mu$ for the case of Fig.~\protect\ref{tb35a1tall:fig}. We
concentrate on the region of parameter space where $\mchi\sim
800\gev$.  The whole marked region is consistent with all the constraints
from colliders and $\bsgamma$. By further imposing the
constraint $0.1<\abundchi<0.2$ one selects only the red region.  Near the
resonance $\mha\simeq2\mchi$, significantly smaller values of
$\sigsip$ become allowed by $0.1<\abundchi<0.2$, (mostly in the
increasingly pure bino region), but become eventually limited from
below, independently of increasing the maximum allowed value of $\mu$.
}
}
\end{center}
\end{figure}

In the same window and in the lower right window we explore the
existence of the {\em lower} limit on $\sigsip$ and its
dependence on the assumed {\em
upper} limit on $\mha$ and $\mu$, respectively. As we can see, the
lowest values of $\sigsip$ are often to a large extent determined by a
somewhat subjective restrictions from above on these parameters.
As one allows either $\mu$ or $\mha$ above $1\tev$ the lower limit on
$\sigsip$ relaxes considerably. 

However, we argue that, by requiring sizable enough $\abundchi$
(\eg, $\abundchi>0.1$), it is possible to set a {\em
parameter--independent} lower bound on $\sigsip$ for a considerable
range of large $\mchi\lsim1020\gev$ (marked with a thick solid line
and left--pointing arrow in Fig.~\ref{tb35a1tdep:fig}). The limit
holds for an arbitrary case of the neutralino (gaugino, higgsino or
mixed), independently of how large $\mu$ and other SUSY parameters are
taken. Let us first consider the gaugino limit $\mu\gg \mtwo$. In
this case the origin of the parameter--independent bound is displayed
in Fig.~\ref{masigtb35a1k:fig} where we plot $\sigsip$ as a function
of the pseudoscalar Higgs $\mha$ for $\mchi\simeq800\gev$ and scan
over all the other parameters.  As $\mha$ and other parameters are
varied, large ranges of $\sigsip$ remain allowed by collider and
indirect constraints, depending on the maximum allowed value of $\mu$.
For each fixed $\mu$, $\sigsip$ decreases proportionally to the fourth
power of $\mhh\simeq\mha$ because of the (typically dominant)
$t$--channel exchange of the heavy Higgs, as the marked cases of $\mu$
in Fig.~\ref{masigtb35a1k:fig} clearly demonstrate. As $\mu$
increases, at fixed $\mchi$ one moves deeper into the gaugino region
and typically finds large $\abundchi>0.2$. By imposing
$0.1<\abundchi<0.2$ one selects only a narrow red (dark) range with a
large bino component. The bino purity ($p_{\bino}=N_{11}^2$) increases
with increasing $\mu$, which normally quickly gives too large
$\abundchi$.  However, for each $\mchi$ one can choose
$\mha\simeq2\mchi$ (roughly $\mha=1600\gev$ in
Fig.~\ref{masigtb35a1k:fig}) in which case $\abundchi$ is reduced to
an allowed level by a wide resonance due to $A$--exchange. This leads
to allowing a much reduced $\sigsip$, while still being consistent
with $0.1<\abundchi<0.2$. However, because of the finite width of the
$A$--resonance, at large enough $\mu$ one reaches the {\em lowest}
value of $\sigsip$ (in this case $2\times 10^{-11}\pb$) which is {\em
parameter--independent}.  Away from the resonance (for example, if one
imposed $\mha<1\tev$), one would obtain the lower bound $\sigsip\gsim
3\times 10^{-8}\pb$, basically independently of whether $\mu<1\tev$ is
imposed or not. In fact, one can see this effect in the lower left
window of Fig.~\ref{tb35a1tdep:fig} where imposing $\mha<1\tev$ causes
the lower limit on $\sigsip$ to suddenly jump up at around
$\mchi\simeq \mha/2\simeq500\gev$.

The above parameter--independent lower limit arises in the gaugino
case and applies to arbitrarily large $\mchi$. The case of the
gaugino--like LSP can be argued to be more attractive as being less
fine--tuned than the higgsino--like one~\cite{chiasdm}. In the MSSM
the LSP is a nearly--pure higgsino for $\mtwo\gsim300\gev$ which, by
applying the assumption of gaugino mass--unification (see
below~(\ref{chidef:eq})) to the gluino mass, implies
$\mgluino=\frac{\alpha_s}{\alpha_2} \mtwo\gsim 1\tev$ and therefore
large and less ``natural'' soft SUSY--breaking scale. Nevertheless, in
the spirit of generality, we need to extend the analysis to the case
of the higgsino--like neutralino.

For the higgsino--like LSP ($\mtwo\gg\mu\simeq\mchi$), if we remain
within the ranges of parameters given in~(\ref{parrange:eq}), we find
$\sigsip$ some two orders of magnitude larger than in the gaugino case
presented above. However, one can in principle reduce $\sigsip$ to
arbitrarily small values by going to the limit of pure enough higgsino
($\mtwo$ in the multi-\tev\ range), in which case the
neutralino--Higgs coupling would be arbitrarily reduced, and by
further suppressing the squark contribution by making them extremely
heavy. It is therefore reasonable to question the existence of the
lower bound on $\sigsip$.  However, in the multi--hundred~\gev\ range
($m_t<\mchi\lsim1020\gev$) $\abundchi$ remains typically very small
$\abundchi\ll 0.1$ although it does increase with $\mchi$. For
$\tanb=35$ it reaches $0.1$ for $\mchi\lsim1020\gev$, $0.2$ for
$\mchi\lsim1.6\tev$ and $0.3$ for $\mchi\lsim2.5\tev$. (These values
decrease somewhat with increasing $\tanb$.) Since, as mentioned above,
we impose $0.1<\abundchi$ (or a similar sizable lower limit on
$\abundchi$), in the range of $\mchi<1\tev$ displayed in
Figs.~\ref{tb35a1tall:fig} and~\ref{tb35a1tdep:fig} the
parameter--independent lower limit on $\sigsip$ holds. (In order to
be clear that the parameter--independent lower bound on $\sigsip$
applies to a general neutralino so long as $\mchi\lsim1020\gev$, in
Figs.~\ref{tb35a1tall:fig} and~\ref{tb35a1tdep:fig}  we have put a
left--pointing arrow at $\mchi=1\tev$.)

For the higgsino--like LSP in the mass range
$1020\gev\lsim\mchi\lsim1600\gev$ the relic abundance is
$0.1<\abundchi<0.2$, basically indpendently of how large $\mtwo$ is,
and no lower bound on $\sigsip$ can in principle be set. The range of
$\mchi$ increases to $2.5\tev$ if we allow $\abundchi<0.3$. Larger
values of $\mchi$ for an arbitrary neutralino gaugino/higgsino
composition are inconsistent with $\abundchi<0.3$. This upper limit on
$\mchi$ relaxes to $2.6\tev$ for $\tanb=10$.

On the other side, at low enough $\mchi$ coannihilation with sleptons
prevents one from deriving a firm lower limit on $\sigsip$. Indeed, by
suitably choosing the slepton mass not too much above $\mchi$, we can
always reduce $\abundchi$ below $0.2$. Thus, it is possible to set
firm lower limits on $\sigsip$ but only for large enough $\mchi$ and
even though they correspond to extremely large values of $\mu$ and
accordingly involve much fine--tunning.

The above discussion of the parameter--independent lower limit on
$\sigsip$ and upper limit on $\mchi$ has been presented in the case of
$\mu>0$ but it obviously applies also to the case $\mu<0$.

The dependence of $\sigsip$ on $\at$ and $\ab$ is rather weak. As
the tri--linear terms deviate from zero, the mass of the lightest
Higgs generally increases due to somewhat larger mass splittings
among the stops and sbottoms. As a result, the normally subdominant
contribution to $\sigsip$ from the $t$--channel $\hl$--exchange is
slightly reduced. 

A far more important effect is that the number of SUSY configurations
satisfying all experimental constraints, especially that from
$\bsgamma$, decreases significantly.  This is because the cancellation
between charged Higgs loop and chargino loop contribution to the
$\bsgamma$ decay rate becomes more inefficient as $\at$ decreases.
For example, for $\at=-1\tev$ only a handful of points remain
allowed. Generally, we have concluded that the regions allowed by
lower values of $\at$ and $\ab$ fall into the regions allowed by the
choice $\at=1\tev$.

\begin{figure}[t!]
\vspace*{-0.75in}
\hspace*{-.70in}
\begin{center}
\begin{minipage}{6.5in}
\epsfig{file=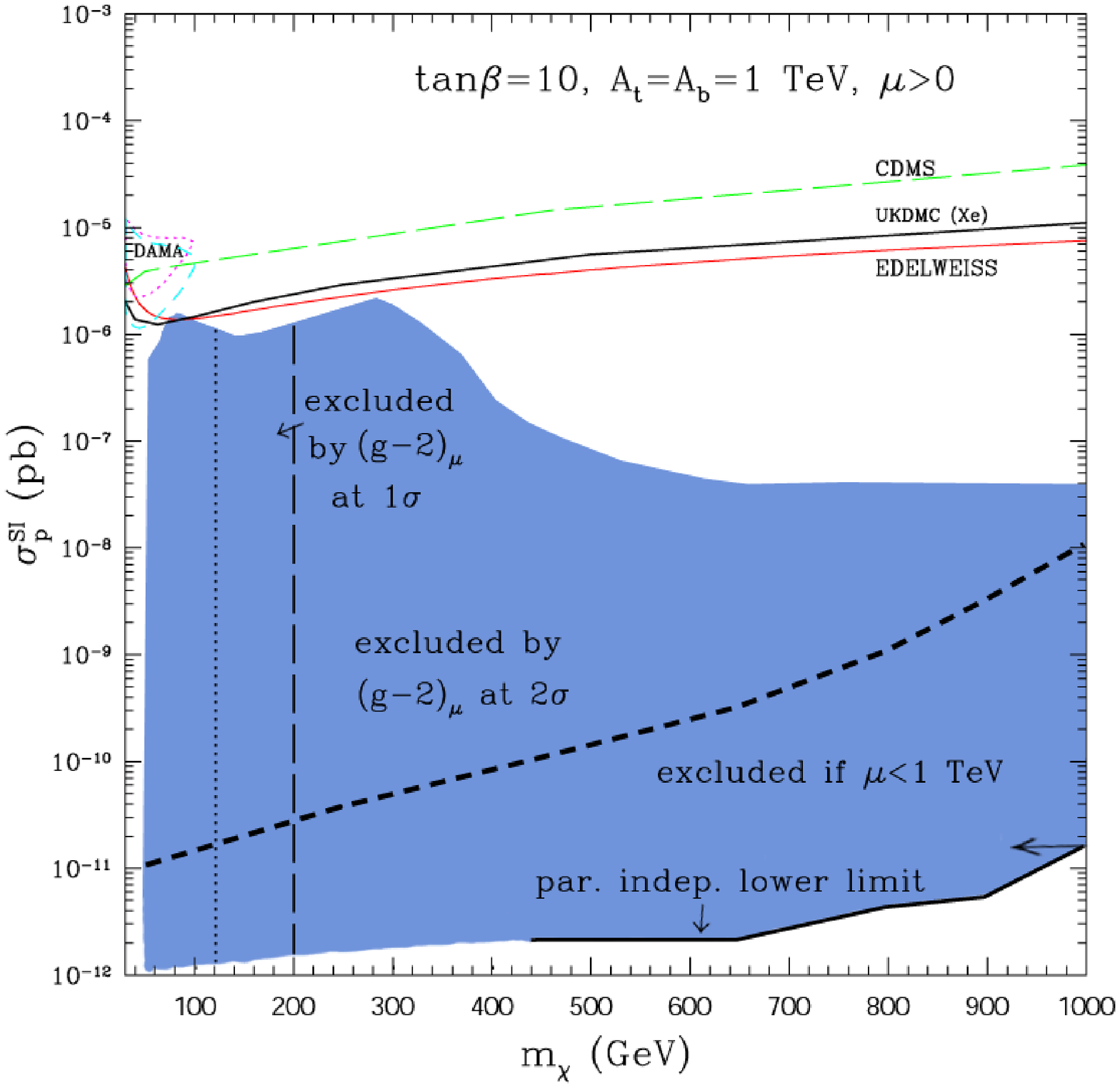,height=3.15in}
\hspace*{-0.18in}
\epsfig{file=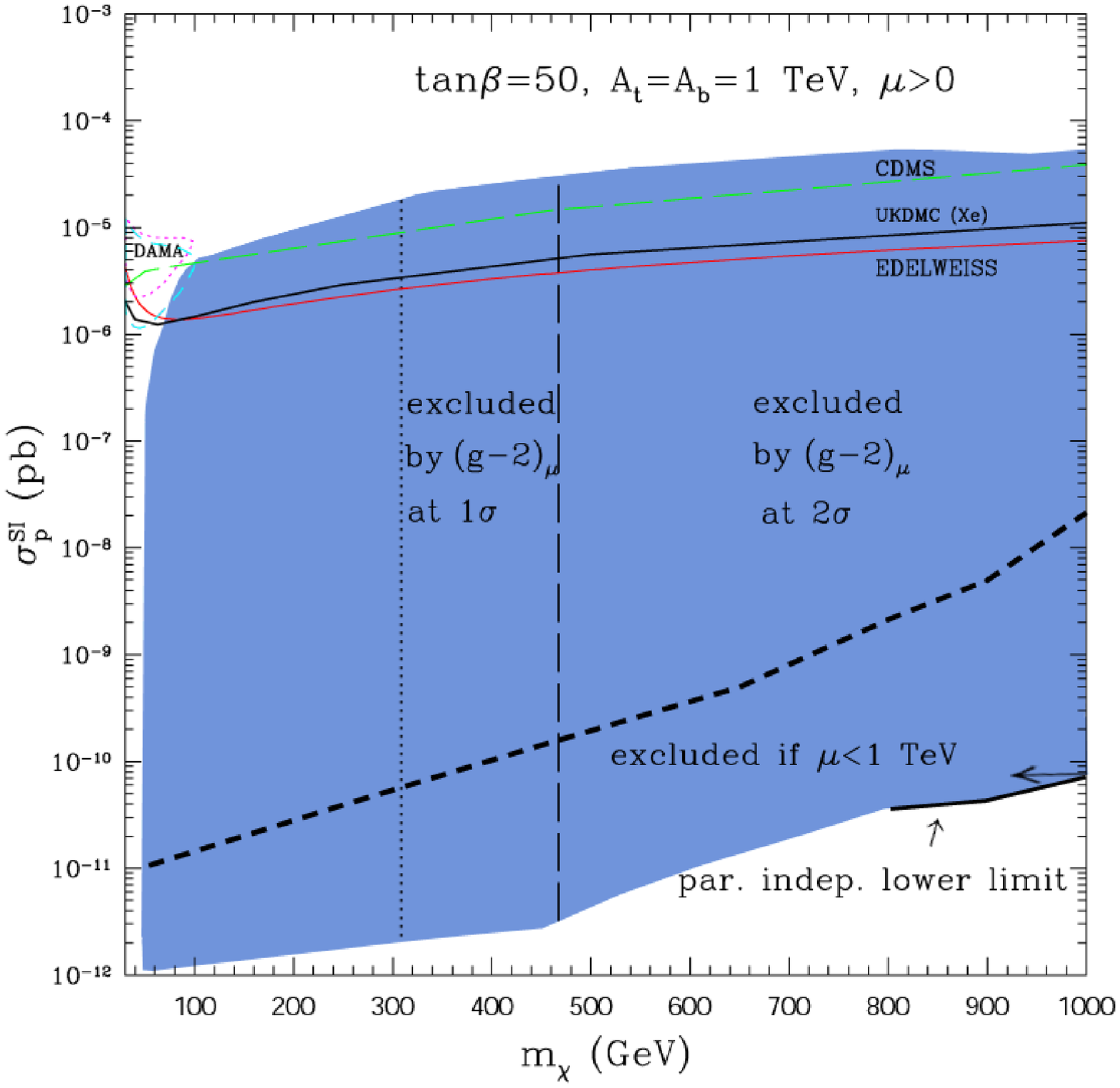,height=3.15in}
\end{minipage}
\caption{\label{tb1050:fig} {\small The same as in
Fig.~\protect\ref{tb35a1tall:fig} but for $\tanb=10$ (left window) 
and 50 (right window). Also marked is the effect of imposing $\mu<1\tev$.
}}
\end{center}
\end{figure}

\begin{figure}[t!]
\vspace*{-0.75in}
\hspace*{-.70in}
\begin{center}
\begin{minipage}{5in}
\epsfig{file=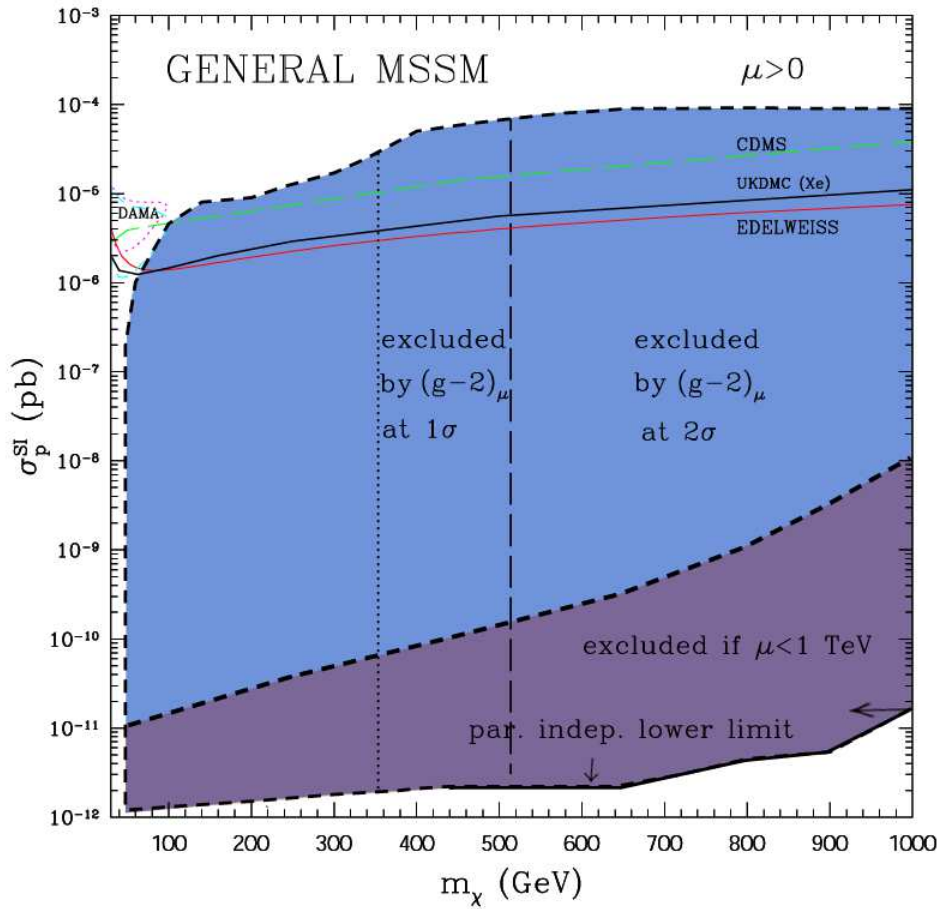,height=5in}
\end{minipage}
\caption{\label{tballmup:fig} {\small 
Ranges of $\sigsip$ in the general MSSM
\vs\ $\mchi$ for $\mu>0$, which are allowed by
collider bounds, $\bsgamma$ and $0.1<\abundchi<0.2$. Also marked are
some results of recent experimental WIMP searches. The thick black
line and a left--pointing arrow 
indicate a {\em parameter--independent} lower bound. The region below
the dashed line is excluded if one imposes the constraint $\mu<1\tev$.
The ranges of
$\mchi$ to the vertical lines are excluded at $1\,\sigma$ and
$2\,\sigma$~CL by the current discrepancy
between the experimental value of $\gmtwo$ and the Standard Model
prediction.
}
}
\end{center}
\end{figure}

In order to display the dependence on $\tanb$, in the left and right
window of Fig.~\ref{tb1050:fig} we present the cases of $\tanb=10$ and
$50$, respectively. Note that, for small $\tanb=10$ the largest
allowed values of $\sigsip$ are roughly an order of magnitude smaller
than at $\tanb=50$ because of the $\tanb$--dependence of the heavy
scalar coupling to down--type quarks, as mentioned earlier. Notice a
significant decrease in the upper ranges of $\sigsip$ at large $\mchi$
for $\tanb=10$, which is caused by exceeding the upper limit
($4.08\times 10^{-4}$) of the allowed range of $\br(B \to
X_s\gamma)$. It is clear that the constraint is more severe in the
case $\tanb =10$ rather than at larger $\tanb$. This may sound
somewhat counter--intuitive since, for example, in the Constrained
MSSM, the constraint from $\bsgamma$ on the (CMSSM) parameter space
becomes more pronounced at larger $\tanb$. This is because, in the
CMSSM the pseudoscalar Higgs mass, hence also the charged Higgs mass,
is typically large, and the corresponding charged Higgs--top quark
loop contribution to $\brbsgamma$ becomes suppressed.  At smaller
$\mhalf$ (thus also $\mchi$ and $\mcharone$) and $\mzero$ (thus also
light enough stop) the $\tanb$--dependent (negative) contribution from
the chargino-stop loop gives too small $\br(B \to X_s\gamma)$, below
the lower experimental limit, thus producing a strong lower bound on
$\mchi$ at not too large $\mzero$. In contrast, in the general MSSM
case, we can choose small values of $\mha$ which is a free parameter.
This small $\mha$ implies a big positive charged Higgs loop
contribution to $\brbsgamma$. For smaller $\tanb$ and at smaller
$\mchi$ this is reduced to an acceptable range by the chargino-stop
loop contribution. However, at large $\mchi$, the chargino-stop loop
cannot cancel the charged Higgs contribution anymore and one exceeds
the upper experimental limit on $\brbsgamma$. Since the chargino loop
contribution is proportional to $\tanb$, at large $\tanb$ the
cancellation can be achieved even at large $\mchi$. For this reason,
in the right window of Fig.~\ref{tb1050:fig} there is no analogous
decrease in the largest allowed $\sigsip$ at large $\mchi$, in
contrast to the left window. Clearly, the constraint from $\bsgamma$
is more severe for smaller $\tanb$.

A similar effect of can be observed in the case of the
$\gmtwo$ constraint. By comparing Fig.~\ref{tb1050:fig} with the upper
right window of Fig.~\ref{tb35a1tdep:fig} one can see that it produces
as stronger upper bound on $\mchi$ at smaller $\tanb$. 
For example, imposing a $1\,\sigma$ bound implies
$\mchi\lsim120\gev$ for $\tanb=10$, while for $\tanb=50$ the bound
moves up to $\mchi\lsim310\gev$, as denoted in the two windows
of Fig.~\ref{tb1050:fig}. Also marked is the effect of imposing
$\mu<1\tev$. While not being a firm constraint, it does, in our
opinion, indicate the region which may be considered as somewhat less
fine--tuned.

As discussed above, for a considerable range of large
$\mchi\lsim1020\gev$ (marked with a thick solid line and
left--pointing arrow in Fig.~\ref{tb1050:fig}), we can again set up the
parameter--independent lower limit on $\sigsip$, in analogy with the
case $\tanb=35$ of Fig.~\ref{tb35a1tdep:fig}. In order to do so, we
had to explore extremely large ranges of $\mu$ up to some $4\tev$ at
smaller $\tanb$ in order to saturate the bound $\abundchi<0.2$.  On
the other hand, at smaller $\mchi\lsim440 (800)\gev$ for $\tanb=10
(50)$ the lower limit on $\sigsip$ remains basically independent of
$\tanb$ since $\abundchi$ at lower $\mchi$ is determined mostly by the
coannihilation with sleptons, as discussed in detail in the case
$\tanb=35$. In this case the lower limit on $\sigsip$ arises from
restricting $\mu$ below the value for which the parameter--independent
lower limit arises at larger $\mchi$. Notice that the thick
line extends to lower $\mchi$ at smaller $\tanb$ because the
efficiency of the neutralino--slepton coannihilation increases with
$\tanb$.

Finally, in Fig.~\ref{tballmup:fig} we summarize the results for the
full scan conducted so far for $\tanb=5,10,35,50,55,60,65$ and
for $\mu>0$. We repeat that, in determining the allowed (blue) region
we applied the constraints from collider searches, $\bsgamma$ and
$0.1<\abundchi<0.2$. The lower limit on $\sigsip$ mostly comes from
low $\tanb$ and, where applicable, we mark with the solid line where
it is parameter--independent. The effect of restricting $\mu<1\tev$ is
marked with a dashed line. Also indicated is the impact of the new
measurement of $\gmtwo$ on the mass of the neutralino. If confirmed,
the $\epem$--based range~(\ref{deltaamu:ref}) will imply rather
stringent upper limits on $\mchi$ 
\bea
\mchi&\lsim&350\gev~~~(1\,\sigma~{\rm CL})\\
\mchi&\lsim&510\gev~~~(2\,\sigma~{\rm CL}).  
\eea
If the $\tau$--based numbers~(\ref{deltaamutau:ref}) are applied instead,
one obtains $\mchi\lsim800\gev~(1\,\sigma~{\rm CL})$ and no upper bound at
$2\,\sigma~{\rm CL}$ (for $\mu>0$).

The vast ranges of $\sigsip$ predicted in
the framework of the general MSSM may be somewhat discouraging to DM
WIMP hunters. It is worth noting, however, that it is the region of
smaller $\mchi$, below a few hundred~$\gev$, that not only is implied
by the new result for $\gmtwo$, but is also theoretically
more favored as corresponding to less fine--tuning. Furthermore,
ranges of very small $10^{-12}\pb\lsim\sigsip\lsim10^{-8}\pb$
generally correspond either to very large (and therefore perhaps
somewhat less natural) values of $\mu$ and/or $\mha$, or become
allowed by selecting slepton masses on the light side, and in the
$\chi$--slepton coannihilation region, within some $20\gev$ of
$\mchi$, which again can be be considered as a finely--tuned case.

At the end, we comment again on the case of the Constrained
MSSM. Because the model is much more restrictive, the ranges of
$\sigsip$ that one obtains in the parameter space allowed by all
constraints, are very much narrower~\cite{efo00,cmssm:recent,
rrn1,lr-latalk,cn0208}. They are also typically somewhat lower than
the largest ones allowed in the general MSSM. For example, at
$\tanb=50$ we find $\sigsip\sim10^{-7}\pb$ at $\mchi=100\gev$ and
$\sigsip\sim 7\times 10^{-11}\pb$ at the largest (neglecting $\gmtwo$)
allowed value of $\mchi=800\gev$.  On the other hand, because the
model is defined at the grand--unified, and not electroweak, scale, in
the case of large $\tanb\gsim50$ and/or large scalar masses,
theoretical uncertainties involved in the running of parameters are
substantial and have much impact on the resulting ranges of both
$\mchi$ and $\sigsip$~\cite{lr-latalk}. We will explore the case of
the Constrained MSSM in an oncoming publication.

\section{Conclusions}\label{conclusions:sec}
We have delineated the ranges of the SI cross section $\sigsip$ in the
general MSSM, which are consistent with  current experimental
bounds and for which one finds the expected amount of dark matter. We
have further discussed the dependence of our results on the 
experimental constraints and on the underlying
theoretical assumptions. While the ranges which we have obtaine extend
over more than six orders of magnitude, we find it encouraging that
the experimental sensitivity that has recently been reached, now
allows one to explore our theoretical predictions for the MSSM. As we
have argued above, smaller values of the WIMP mass and also larger
values of $\sigsip$ may be considered as more natural, which will
hopefully be confirmed by a measuring a positive WIMP detection signal
in the near future.

\bigskip

\acknowledgments 
We thank M.~Brhlik for his help in testing our
numerical code for computing the cross section for WIMP--proton
scattering. L.R. is grateful to A.~Czarnecki for helpful comments
about theoretical and experimental aspects of $\gmtwo$. This work was
supported in part by the EU Fifth Framework network "Supersymmetry and the
Early Universe" (HPRN-CT-2000-00152).

\end{document}